\documentclass[aps,prb,reprint,floatfix,superscriptaddress]{revtex4-1}
\usepackage[english]{babel}                                                           
\usepackage[utf8]{inputenc}                                                            
\usepackage[T1]{fontenc}
\usepackage{bbold}
\usepackage{epsfig}
\usepackage{amsmath}
\usepackage{hyphenat}
\usepackage{units}
\usepackage{hyperref}
\usepackage{placeins}
\usepackage{enumerate}
\usepackage[justification=raggedright]{caption}	
\usepackage[colorinlistoftodos]{todonotes}			  				

\usepackage{lipsum}

\begin{document}

\title{Modifications of the Lifshitz-Kosevich formula in two-dimensional Dirac systems}
\author{Carolin K\"uppersbusch}
\affiliation{Institute for Theoretical Physics and Center for Extreme Matter and Emergent Phenomena,
Utrecht University, Leuvenlaan 4, 3584 CE Utrecht, The Netherlands}
\affiliation{Institut für theoretische Physik, Universität zu Köln, Zülpicher Straße 77, 50937 Köln, Germany}
\author{Lars Fritz}
\affiliation{Institute for Theoretical Physics and Center for Extreme Matter and Emergent Phenomena,
Utrecht University, Leuvenlaan 4, 3584 CE Utrecht, The Netherlands}

\begin{abstract}
%We derive a full quantitative formula which describes the amplitude and frequency of magnetic oscillations in two-dimensional Dirac systems. %The investigations are on the basis of graphene, but they generally also hold for other two-dimensional Dirac systems.
%, such as the surface of a three-dimensional topological insulator.
Starting from the Luttinger-Ward functional \cite{Luttinger1960} we derive an expression for the oscillatory part of the grand potential of a two dimensional Dirac system in a magnetic field. We perform the computation for the clean and the disordered system, and we study the effect of electron-electron interactions on the oscillations. Unlike in the two dimensional electron gas (2DEG), a finite temperature and impurity scattering also affects the oscillation frequency. Furthermore, we find that in graphene, compared to the 2DEG, additional interaction induced damping effects occur: to two-loop order electron-electron interactions do lead to an additional damping factor in the amplitude of the Lifshitz-Kosevich(LK)-formula. \\

\end{abstract}
\maketitle

\section{Introduction}
A major breakthrough in the investigations of electron-electron interactions in many-body systems was the Landau theory of the Fermi liquid \cite{landau1957}. It explains why a system of strongly interacting particles can be described by a system of non-interacting quasiparticles, which allows for simple theoretical models to describe phenomena in condensed matter system. However, in Dirac systems, Fermi liquid theory is not straightforwardly applicable \cite{Kotov2012}. It has been shown that they can be described by a marginal Fermi liquid \cite{Gonzales1999}: since the screening length diverges at the Dirac point, electron-electron interactions are in principle expected to play a significant role. \\
One important example of a two-dimensional Dirac system is graphene. 
In general there exist two different predictions on the effect of Coulomb-interactions in graphene. Assuming a weak coupling perspective, electron-electron interactions are assumed to renormalize the Fermi velocity according to $v_F \rightarrow v_F\ln \frac{\Lambda}{k}$, where $k$ is the momentum and $\Lambda$ is a high-energy cutoff\cite{Gonzales1994, Elias2011, Yu2013}.  
At strong coupling one expects the system to be an excitonic insulator \cite{Khveshchenko,Drut2009}. Different approaches yield different critical interaction strengths; some suggest that graphene in vacuum might be an insulator while experimental evidence rather points towards it being weakly coupled\cite{Drut2013, Elias2011}.\\
There are only very few experiments which allow to deduce information about electron-electron interactions and it would be desirable to have other experimental probes which one can compare to theory predicitions. 
%There are only very few experiments which allow to deduce information about the electron-electron interactions in a system. 
A standard experiment in the determination of electronic properties of conductors is the measurement of quantum oscillations in both transport (Shubnikov-de Haas) and thermodynamic (de Haas-van Alphen) quantities. Generally, the amplitude of the oscillations is described by the Lifshitz-Kosevich (LK) formula\cite{LifshitzKosevich,Luttinger1960,Shoenberg1984}. As electron-electron interactions enter the oscillation amplitude, one can extract information about it by fitting the LK-formula to the measured amplitude\cite{Elias2011, Shoenberg1984}. In a two-dimensional electron gas (2DEG) in the Fermi liquid regime, the amplitude of the LK-formula was shown to {\it not} contain an additional damping factor due to electron-electron interactions which could be associated with an effective temperature, but instead electron-electron interactions only affect the oscillations by renormalizing the cyclotron frequency or fermionic mass. \cite{Mirlin2006, Martin2003} It has to be pointed out that a similar statement was made for the electron-phonon problem~\cite{Shoenberg1984,FowlerPrange,EngelsbergSimpson}.\\
Since a Dirac systems near the charge neutrality point is not a true Fermi liquid \cite{Kotov2012, Gonzales1999}, we ask the question whether the LK-formula for a Fermi liquid still holds. We also ask this question in view of experiments where the Fermi liquid LK-formula was used to extract interaction effects from the damping of the amplitude of magnetic oscillations in graphene\cite{Elias2011}, albeit for transport measurements. 

\subsection{Model}

Dirac systems are characterized by a linear dispersion resulting from the Hamiltonian,
%It is well known that in contrast to a standard two dimensional electron gas (2DEG) the effective low-energy description in graphene assumes the form of a Dirac theory
\begin{equation}
\mathcal{H}=v_F k_x \hat{\sigma}_x+ v_F k_y \hat{\sigma}_y-\mu\; \mathbb{1} \;
\end{equation}
where $\mathcal{H}$ is the effective low-energy Bloch Hamiltonian, $v_F$ is the Fermi velocity, $\mu$ corresponds to the chemical potential and the Pauli matrices $\hat{\sigma}_x$ and $\hat{\sigma}_y$ in graphene for instance act in sublattice space while on the surface of a three dimensional topological insulator they would act in spin space.

The orbital effect of an external magnetic field, $B$, is in this formulation accounted for in the standard form of the minimal coupling according to $\vec{k} \to \vec{k}-e \vec{A}$, where $\vec{A}$ denotes the gauge field and $e$ is the electron charge. In a magnetic field the spectrum for a single Dirac cone is given by
\begin{align}
E_{\pm}(m)&=\pm \omega_c \sqrt{m}\\  \textup{with}\quad
\label{omegac}
\omega_c&=v_F\sqrt{2 e B}
\end{align}
where $m$ is the Landau level index. These Landau levels are highly degenerate and interestingly the zeroth Landau level only lives in one component of the spinor. 
%We expect that for a large chemical potential $\mu$ compared to all other energy scales such as temperature $T$ and cyclotron frequency $\omega_c$ the system behaves very similar to a standard 2DEG and consequently the effect of magnetooscialltions should be described by Eq.~\eqref{eq:omega} and Eq.~\eqref{eq:LK}.

\subsection{Main questions and summary of the results}

The purpose of this paper is to investigate the applicability of the standard LK formula upon approaching the Dirac point. There are three aspects to this question which we investigate separately in Section \ref{LKformula}: 
\begin{itemize}
\item (I) What is the effect of temperature as the temperature approaches the value of the chemical potential in graphene? 

\item (II) How is this modified in the presence of disorder? 

\item (III) Is the effect of inelastic processes also to just modify the effective mass or do we have an additional damping term due to interaction effects associated with another inelastic Dingle damping temperature? 
\end{itemize}
The main results of the paper regarding these questions are:

(I) Temperature acts in a similar manner as in standard two-dimensional electron gases. However, there is an interesting modification upon approaching the Dirac point, which is, that temperature modifies the oscillation frequency, meaning the oscillation frequency is not a pure geometrical quantity any more. Temperature also provides a cutoff for the quantum oscillations meaning that as soon as temperature is on the order of the chemical potential the quantum oscillations die altogether.  

(II) The effect of disorder can be described in the same way as in the two-dimensional electron gas, accounted for by a Dingle temparature. In Dirac systems, however, there also is a damping term due to the coupling of temperature and disorder. Furthermore, disorder in graphene also affects the oscillation frequency, which is not the case in a 2DEG. 

(III) Unlike in a 2DEG, electron-electron interactions in graphene lead to renormalization effects {\it and} inelastic effects.

We performed all our calculations for the specific case of graphene, but the results are also applicable to other Dirac systems without any restrictions. The only specification is a factor of four which is a consequence of the spin and valley degeneracy in graphene.
%%%%%%%%%%%%%%%%%%%%%%%%%%%%%%%%%%%%
% The oscillatory grand potential
%%%%%%%%%%%%%%%%%%%%%%%%%%%%%%%%%%%%

\section{The oscillatory grand potential}
\label{LKformula}
Our starting point is the Luttinger-Ward functional \cite{Luttinger1960} which relates the thermodynamic potential $\Omega$ of the system to its Green function, $\hat{G}$,
\begin{align}
\Omega=-T \textup{Tr ln}(-\hat{G}^{-1})-T\textup{Tr}(\hat{G}\hat{\Sigma})+\Omega'.
\end{align}
The Green function for graphene in a magnetic field reads
\begin{equation}
\label{Greenfunctiongraphenemagnfield}
\hat{G}^{-1}_m(i\omega_n)=\left( \begin{array}{cc} i\omega_n+\mu & \omega_c \sqrt{m} \\ \omega_c \sqrt{m} & i\omega_n+\mu \end{array} \right)-\hat{\Sigma}(i\omega_n,m). \nonumber \\
\end{equation}
(We set the Boltzmann constant, $k_B\equiv 1$, as well as Planck's constant, $\hbar\equiv 1.$)
The trace implies summation over the Landau level index $m$, the fermonic Matsubara frequencies $\omega_n=\pi T(2n+1)$, and the different degenerate states within one Landau level. The self-energy $\hat{\Sigma}$ accounts for disorder ($\hat{\Sigma}_{dis}$) or electron-electron interactions $(\hat{\Sigma}_{ee})$. The terms $T\textup{Tr}(\hat{G}\hat{\Sigma})$ and $\Omega'$ are introduced to avoid overcounting of diagrams. Their oscillatory parts cancel each other \cite{Mirlin2006} such that the magnetic oscillations are fully described by
\begin{align}
\label{LuttingerWard}
\Omega_{mo}=-T \textup{Tr ln}(-\hat{G}^{-1})=-DT\sum_m\sum_{\omega_n}\ln(-g^{-1}_{m}(i\omega_n)),
\end{align}
where $g_m^{-1}$ are the eigenvalues of the matrix $\hat{G}^{-1}$. The factor $D$ accounts for the sum over degenerate Landau levels,
\begin{equation}
\label{degeneracyfactorLLgraphene}
D=\frac{e B L^2}{\hbar\pi}=\frac{\omega_c^2L^2}{2\pi v_F^2},
\end{equation}
where $L$ is the size of the system. 
We use the Poisson summation formula which relates the summation of a function to the function's continuous integral:
\begin{align}
\label{Poisson}
\sum_{m=0}^\infty f_m&=\lim_{\epsilon\rightarrow0^+}\sum_{m=-\infty}^\infty \int_{-\epsilon}^\infty  {\rm d}x f(x) \delta(x-m)\nonumber\\
&=\sum_{l=-\infty}^\infty \int_{0}^\infty  {\rm d}x f(x) e^{i2\pi l x}\nonumber\\
&=\int_{0}^\infty  {\rm d}x f(x) +2\sum_{l=1}^\infty \int_{0}^\infty  {\rm d}x f(x) \cos{2\pi l x}
\end{align}
%Philosophically this transformation means that we go from discrete Landau levels to a continuous picture, where a constant density of states is modulated by oscillating modes, see Figure \ref{fig:poissonformula}. 
This approach is advantageous when the Landau levels are sufficiently broadened due to e.g. disorder or temperature which we assume throughout this paper. In this case we obtain a small parameter, $\omega_c/\alpha_{\textup{dis}}$ ($\alpha_{\textup{dis}}$ is the disorder potential) and $\omega_c/T$, respectively, such that the series can be truncated.\\ 
The first term describes the $B=0$ state and consequently does not contribute to the oscillations. We insert the function defined in Eq.~\eqref{LuttingerWard} into Eq.~\eqref{Poisson} and integrate by parts,
\begin{align}
\label{genericoscintegral}
\Omega_{mo'}=&-2DT\sum_{\omega_n}\sum_{l=1}^\infty \int_{-\epsilon}^\infty  {\rm d}x \ln(-g^{-1}(x,i\omega_n)) \cos{2\pi l x}\nonumber\\
=&-2DT\sum_{\omega_n}\sum_{l=1}^\infty \left[\ln[-g^{-1}(x,i\omega_n)]\frac{\sin(2\pi lx)}{2\pi l}\right]_0^\infty\nonumber\\+&2DT\sum_{\omega_n}\sum_{l=0}^\infty\int_0^\infty \frac{1}{-g^{-1}(x,i\omega_n)}\frac{\textup{d}}{\textup{d}x}\Big(-g^{-1}(x,i\omega_n)\Big)\nonumber\\&\quad\quad\quad\quad\quad\quad\times\frac{\sin(2\pi lx)}{2\pi l}\textup{d}x.
\end{align}
The first term is non-oscillatory and finite due to a cut-off in the Green function. The oscillatory part consequently reads
\begin{align}
\tilde{\Omega}&=2DT\sum_{\omega_n}\sum_{l=0}^\infty\int_0^\infty \frac{1}{-g^{-1}(x,i\omega_n)}\frac{\textup{d}}{\textup{d}x}\Big(-g^{-1}(x,i\omega_n)\Big)\nonumber\\&\quad\quad\quad\quad\quad\quad\quad\quad\times\frac{\sin(2\pi lx)}{2\pi l}\textup{d}x.
\end{align}
With the following ansatz for the Green function
\begin{eqnarray}
\hat{G}^{-1}&=&\left(\begin{array}{cc}
g_1 & g_2\\
g_2&  g_1 \end{array}\right),
\end{eqnarray}
which has eigenvalues
\begin{align}
g^{-1}=g_1\pm |g_2|,
\end{align}
we find
\begin{align}
\label{genericoscintegral}
\tilde{\Omega}=\frac{DT}{\pi}\sum_{\omega_n}\sum_{l=1}^\infty\frac{1}{l}\int_0^\infty \left[\frac{\sin(2\pi lx)\frac{d}{dx}(g_1^2-g_2^2)}{g_1^2-g_2^2}\right]\textup{d}x.
\end{align}
In the following sections, Sec. \ref{cleanlimit} - Sec. \ref{eeWW}, we specify this generic expression according to the Green function of the discussed system. 

%%%%%%%%%%%%%%%%%%%%%%%%%%%%%%%%%%%%
% The clean limit
%%%%%%%%%%%%%%%%%%%%%%%%%%%%%%%%%%%%

\subsection{The clean limit}
\label{cleanlimit} 
In the limit of a clean system the Green function is simply the free Green function, meaning we have
\begin{equation}
\label{G0graphenemagnfield}
\hat{G}_m^{0\;-1}(i\omega_n)=\left( \begin{array}{cc} i\omega_n+\mu & \omega_c \sqrt{m} \\ \omega_c \sqrt{m} & i\omega_n+\mu \end{array} \right) \;.
\end{equation}
Inserting this Green function into the oscillatory potential, Eq.~\eqref{genericoscintegral}, we find
\begin{align}
\label{omegacleanbeforerestheorem}
\tilde{\Omega}=-\frac{\omega_c^2DT}{\pi}\sum_{\omega_n}\sum_{l=1}^\infty\frac{1}{l}\int_0^\infty \left[\frac{\sin(2\pi lx)}{(i \omega_n+\mu)^2-\omega_c^2x}\right]\textup{d}x.
\end{align}
The integral can be computed using residue theorem and the calculation is performed in Appendix \ref{app:clean}. We obtain,
\begin{equation}
  \tilde{\Omega}_{\rm{osc}}=\frac{2T\omega_c^2L^2}{\pi v_F^2}\sum_{l=1}^\infty \sum_{\omega_n>0}^{|\mu|}\frac{e^{-\frac{4\pi l}{\omega_c^2}\omega_n |\mu|}}{l} \cos \left(\frac{2\pi l \left(\mu^2-\omega_n^2 \right)}{\omega_c^2} \right).
  \label{eq:oscclean}
\end{equation}
\\
We will come back to a discussion of this expression in Sec.~\ref{ch:comparison}.
%%%%%%%%%%%%%%%%%%%%%%%%%%%%%%%%%%%%
% Disorder
%%%%%%%%%%%%%%%%%%%%%%%%%%%%%%%%%%%%
\subsection{Weak disorder limit}
\label{disorder}
We do not attempt to make a realistic modelling of the properties of graphene with disorder but instead stick to the most simple treatment of disorder within the self-consistent Born approximation along the lines of Ref.~\cite{BriskotMirlin}. The disorder induced self-energy is explicitly computed in Appendix \ref{app:selfenergydis}.
In the limit of weak magnetic field ({\it {i.e.}} not well separated Landau levels) and white noise disorder, $\omega_c\ll \alpha_{\textup{dis}}$ ($\alpha_{\textup{dis}}$ is the strength of the disorder potential), the self-energy can be assumed as
\begin{eqnarray}
\hat{\Sigma}_{\textup{dis}}&= &-\alpha_{\textup{dis}} (i\omega_n-\mu)\ln \left(\frac{v_F^2 \Lambda^2-(i\omega_n-\mu)^2}{-(i\omega_n-\mu)^2}\right)\mathbb{1} \nonumber \\ &\approx&  -\alpha_{\textup{dis}}(i\omega_n-\mu)\ln \left(\frac{v_F^2 \Lambda^2}{-(i\omega_n-\mu)^2}\right) \mathbb{1}\;,
\end{eqnarray}
where $v_F$ is the Fermi velocity and $\Lambda$ is a high energy cutoff. This self-energy is diagonal and independent of energy. Consequently the oscillatory potential, Eq.\eqref{genericoscintegral}, simplifies to 
\begin{equation}
\label{oscintdis}
\tilde{\Omega}=-\frac{\omega_c^2DT}{\pi}\sum_{\omega_n} \sum_{l=1}^{\infty}\frac{1}{l}\int_{0}^\infty  {\rm d}x \frac{\sin(2\pi lx)}{\left(i\omega_n+\mu-\Sigma_{\textup{dis}}\right)^2-\omega_c^2x}.
\end{equation}
 We obtain 
\begin{widetext}
\begin{eqnarray}
\nonumber \tilde{\Omega}_{\rm{osc}}&=&-\frac{2T\omega_c^2L^2}{\pi v_F^2}\sum_{l=1}^{\infty}\frac{1}{l}\sum_{\omega_n>0}^{\frac{|\mu|}{1+\alpha_{\textup{dis}}\frac{\pi}{2}}}e^{-\frac{4\pi l}{\omega_c^2}\left((\mu^2-\omega_n^2)(\pi\alpha_{\textup{dis}}-2\phi\alpha_{\textup{dis}})+|\mu|\omega_n(1+2\alpha_{\textup{dis}} \Gamma)\right)}\\&&\quad\quad\quad\quad\quad\quad\quad\quad\quad\quad\times \cos\left(\frac{2\pi l}{\omega_c^2}\left( (\mu^2-\omega_n^2)(1+2\alpha_{\textup{dis}} \Gamma)+4\alpha_{\textup{dis}}|\mu|\omega_n(2\phi-\pi)\right)\right)
\end{eqnarray}
\end{widetext}
\begin{equation}\label{eq:Gamma}
\textup{with} \  \Gamma=\ln \left(\frac{(v_F \Lambda)^2}{\omega_n^2+\mu^2}\right) \textup{and} \  \phi=\arctan\left(\frac{\omega_n}{|\mu|}\right)\;.
\end{equation}
The concrete computation of the integral is performed in Appendix \ref{app:dis} and a discussion is again given in Sec.~\ref{ch:comparison}.
%%%%%%%%%%%%%%%%%%%%%%%%%%%%%%%%%%%%
% The effect of inelastic scattering on magnetooscillations
%%%%%%%%%%%%%%%%%%%%%%%%%%%%%%%%%%%%

\subsection{The effect of inelastic scattering on magnetooscillations}
\label{eeWW}
Within this section we investigate the effect of interactions in perturbation theory.
We first calculate the oscillatory part of the grand potential with an ansatz for an energy dependent self-energy $\hat{\Sigma}_{ee}$. Then we will compute the interaction induced self-energy for graphene to second order in perturbation theory. We will perform this calculation in $k$-space and for zero chemical potential. This is a strong simplification and we will comment on it later. We will see that for finite temperatures the imaginary part contributes an additional damping factor to the LK-amplitude.\\
We make the following ansatz for the self-energy, which we will motivate below by means of an explicit calculation, whose details can be found in Appendix \ref{app:selfenergyee},
\begin{eqnarray}
\label{ansatzselfenergyeegraphene}
\hat{\Sigma}_{ee}(m,i\omega_n)&=&\left(\begin{array}{cc}
i\omega_n(1-Z^{-1}) & (1-Z_{v_F})\omega_c\sqrt{m}\\
(1-Z_{v_F})\omega_c\sqrt{m}&  i\omega_n(1-Z^{-1}) \end{array}\right)\nonumber \\ &+&\left(\begin{array}{cc}
i\delta\Sigma''(\omega_c\sqrt{m},\omega_n) & \delta\Sigma'(\omega_c\sqrt{m},\omega_n) \\
\delta\Sigma'(\omega_c\sqrt{m},\omega_n) & i\delta\Sigma''(\omega_c\sqrt{m},\omega_n)
\end{array}\right).\nonumber\\
\end{eqnarray}
Here, we assumed that $Z$ and $Z_{v_F}$ account for logarithmic renormalizations and do not explicitly depend on energy. $\delta\Sigma''$ and $\delta\Sigma'$ are real and correspond to non-logarithmic contributions which potentially depend on temperature. The Green function reads,\\
\begin{eqnarray}
\hat{G}^{-1}(m,i\omega_n)&=&\left(\begin{array}{cc}
(i\omega_n  +\mu)Z^{-1} & Z_{v_F}\omega_c\sqrt{m}\\
Z_{v_F}\omega_c\sqrt{m}&  (i\omega_n +\mu)Z^{-1} \end{array}\right)\nonumber \\ &-&\left(\begin{array}{cc}
i\delta\Sigma''(\omega_c\sqrt{m},\omega_n) & \delta\Sigma'(\omega_c\sqrt{m},\omega_n) \\
\delta\Sigma'(\omega_c\sqrt{m},\omega_n) & i\delta\Sigma''(\omega_c\sqrt{m},\omega_n)
\end{array}\right).\nonumber\\
\end{eqnarray}
The eigenvalues of this matrix are
\begin{align}
g_m^{-1}&=(i\omega_n+\mu)Z^{-1}-i\delta\Sigma''\mp(\sqrt{m}\omega_cZ_{v_F}-\delta\Sigma')
\end{align}
We linearize the denominator of Eq. \eqref{genericoscintegral} around the pole $x_0$ with $g_1^2-g_2^2\big|_{x=x_0}=0$,
 \begin{align}
\tilde{\Omega}&=\frac{DT}{\pi}\sum_{\omega_n}\sum_{l=1}^\infty\frac{1}{l}\int_0^\infty \left[\frac{\sin(2\pi lx)\frac{d}{dx}(g_1^2-g_2^2)}{(x-x_0)\frac{d}{dx}(g_1^2-g_2^2)|_{x=x_0}}\right]\textup{d}x.
\end{align}
%\begin{align}
%\Omega&=\frac{DT}{\pi}\sum_{\omega_n}\sum_{l=1}^\infty\frac{1}{l}\sum_{\lambda=\pm1}\int_0^\infty \left[\frac{1}{-i\omega_nZ^{-1}-\mu+i\delta\Sigma''+\lambda\sqrt{x}\omega_cZ_{v_F}-\lambda\delta\Sigma'}\right]\nonumber\\&\times\frac{\lambda\omega_cZ_{v_F}}{2\sqrt{x}}\sin(2\pi lx)\textup{d}x.
%\end{align}
We expand the pole in powers of the interaction parameter $\alpha$, which we will define below ($\delta\Sigma'$ and $\delta\Sigma''$ are quadratic in $\alpha$), and write,
\begin{align*}
x_0\approx x_0^{(0)}+x_0^{(2)}.
\end{align*}
To lowest order we have
\begin{equation}
 \omega_c^2Z_{v_F}^2x_0^{(0)}=\left(i\omega_n+\mu\right)^2Z^{-2}.
 \end{equation}
To quadratic oder we find
\begin{equation}
\omega_c^2Z_{v_F}^2x_0^{(2)}=2i\delta\Sigma''(-i\omega_n-\mu)Z^{-1}+2\delta\Sigma'(x_0^{(0)})\sqrt{x_0^{(0)}}\omega_cZ_{v_F}.
\end{equation}
Computing the integral using residue theorem, we obtain,
\begin{widetext}
\begin{eqnarray}
\label{LKformulaeeWWgraphene}
\tilde{\Omega}_{osc}&=&2DT\sum_{l=1}^\infty\frac{1}{l}\sum_{\omega_n>0}^{\approx|\mu|}e^{-\frac{4\pi l}{\omega_c^2Z_{v_F}^2}\left(\omega_n\mu Z^{-2}+(\Im\{\delta\Sigma''\}+\Re\{\delta\Sigma'\})\omega_nZ^{-1}+(\Im\{\delta\Sigma'\}-\Re\{\delta\Sigma''\})\mu Z^{-1}\right)}\nonumber\\&\times&\cos\left(\frac{2\pi l}{\omega_c^2Z_{v_F}^2}\left((-\omega_n^2+\mu^2)Z^{-2}+2(\Re\{\delta\Sigma''\}-\Im\{\delta\Sigma'\})\omega_nZ^{-1}\right.\right.+2(\Im\{\delta\Sigma''\}+\Re\{\delta\Sigma'\})\mu Z^{-1}\bigg)\Bigg).%\nonumber
\end{eqnarray}
\end{widetext}
In Appendix \ref{app:selfenergyee} we compute the interaction induced self-energy to second order in the fine structure constant $\alpha=e^2/(4 \pi \epsilon_0 \epsilon_r v_F)$ ($\epsilon_0$ being the vaccuum dielectric constant and $\epsilon_r$ the relative permittivity).
We perform the calculation in k-space and not in the Landau-level basis. This approach is questionable for instance in light of the well-known magnetic catalysis in systems with Dirac fermions in a magnetic field~\cite{Gusynin94}. However, we believe that this approach is justified in the limit of sufficient thermal Landau level broadening, $T>\omega_c$, since in that limit the 'effective' density of states which enters the computation is closer to the one without a magnetic field than to the singular one with the Landau levels. To lowest order in $\alpha$ we find
\begin{equation}
\hat{\Sigma}^{(1)}_{ee} ({\bf  k},i \omega_n)=\frac{\alpha v_F{\bf k}\hat{\boldsymbol{\sigma}}}{4 }\ln\left(\frac{4\Lambda}{v_Fk}\right).
\end{equation}
This term has no information about inelastic scattering processes which is why we go to two-loop order. There we find (note that we discard crossed diagrams here and only concentrate on the large-N diagrams)
\begin{eqnarray}
\hat{\Sigma}^{(2)}_{ee} ({\bf  k},i \omega_n)&=&- \frac{\alpha^2}{2}  ( i \omega_n \mathbb{1}+ v_F {\bf{k}} \hat{\boldsymbol\sigma} )\nonumber\\&\times& \left(\frac{1}{6} \ln \left(\frac{\Lambda ^2 v_F^2}{k^2 v_F^2+\omega_n^2}\right)-\frac{1}{3}\ln (2)+\frac{5}{9}\right).\nonumber\\
\end{eqnarray}
Within renormalized perturbation theory we can define the renormalization $Z$-factors we introduced in the ansatz for the self-energy, Eq.\eqref{ansatzselfenergyeegraphene} (note that we use a non-standard definition). We only include the logarithmically dependent parts since the other parts are irrelevant for the flow equations. They read
% \begin{equation}
%&& i\omega_n-v_F {\bf k}\hat{\boldsymbol{\sigma}}}-\frac{\alpha}{4}v_F {\bf k}\hat{\boldsymbol{\sigma}}}\ln \frac{4\Lambda}{v_F k}+\frac{\alpha^2}{8}\left(i\omega_n+v_F {\bf k}\hat{\boldsymbol{\sigma}}}\right )\frac{2}{3} \ln \frac{\Lambda^2}{v_F^2 k^2+\omega_n^2} \nonumber \\ &=&  i\omega_n \left (1+\frac{\alpha^2}{8}\ln \frac{\Lambda^2}{v_F^2 k^2+\omega_n^2}\right)-v_F {\bf k}\hat{\boldsymbol{\sigma}}}\left(1+\frac{\alpha}{4}\ln \frac{4\Lambda}{v_F k}-\frac{\alpha^2}{8}\frac{2}{3} \ln \frac{\Lambda^2}{v_F^2 k^2+\omega_n^2}\right)
%\end{equation}
%meaning we have
\begin{eqnarray}
Z^{-1}&=&1+\frac{\alpha^2}{12}\ln \frac{\Lambda^2v_F^2}{v_F^2 k^2+\omega_n^2}\nonumber \\ Z_{v_F}&=&1+\frac{\alpha}{4}\ln \frac{4\Lambda}{v_F k}-\frac{\alpha^2}{12} \ln \frac{\Lambda^2v_F^2}{v_F^2 k^2+\omega_n^2}\nonumber \\ &+&\frac{\alpha^2}{16}\ln^2 \frac{4 \Lambda}{v_F k}\;.
\end{eqnarray}
While $Z$ corresponds to the field renormalization, as it renormalizes $\omega_n$, $Z_{v_F}$ renormalizes $\omega_c$ and can thus be identified with the renormalization factor of the Fermi velocity (cf. Eq. \eqref{omegac}). This implies that the Green function reads
\begin{equation}
\hat{G}^{-1}({\bf{k}},i\omega_n)=Z^{-1}((i\omega+\mu) \mathbb{1}-v_F {\bf k}\hat{\boldsymbol{\sigma}}Z Z_{v_F})
\end{equation}
and consequently we can define a renormalized Fermi velocity $v_F^R$ as 
\begin{equation}
\label{renormalizedvF}
v_F^R=v_F Z Z_{v_F}
\end{equation}
where $v_F$ is the bare Fermi velocity. Exploiting $\frac{dv_F}{d\ln \Lambda}=0$ we obtain the flow of the renormalized Fermi velocity as
\begin{eqnarray}
\frac{\textup{d} v_F^R}{\textup{d}\ln \Lambda}&=&v_F\left( \frac{\partial Z Z_{v_F}}{\partial\ln \Lambda}+\frac{\partial Z Z_{v_F}}{\partial\alpha}\frac{\partial\alpha}{\partial v_F^R}\frac{\partial v_F^R}{\partial\ln\Lambda} \right)\nonumber \\ &=&\frac{v_F^R}{Z Z_{v_F}}\left( \frac{\partial Z Z_{v_F}}{\partial\ln \Lambda}-\frac{\alpha}{Z Z_{v_F}}\frac{\partial Z Z_{v_F}}{\partial\alpha}\frac{\partial Z Z_{v_F}}{\partial\ln\Lambda} \right)\nonumber \\ &=&v_F^R\left(\frac{\alpha}{4}-\frac{\alpha^2}{3} \right) +\mathcal{O}(\alpha^3)\;.
\end{eqnarray}
The flow of $\alpha$ itself to lowest order in perturbation theory (one loop) is given by $\frac{d\alpha}{d \ln \Lambda}=-\frac{\alpha^2}{4}$. This means renormalizations of $\alpha$ itself do not interfere with this result since they are of higher order in $\alpha$ leading to lowest order contributions at $\mathcal{O}(\alpha^3)$.
The flow equation implies a critical $\alpha_c=3/4$, which potentially describes a repulsive critical point separating weak coupling from strong coupling. However, in a strict large-N limit to all orders in $\alpha$ the absence of such a critical point was shown by Son in Ref. \cite{Gonzales1994,Son2007}, meaning that the critical point most likely is an artefact of the order of approximation.\\
For additional effects of inelastic scattering we have also investigated $\hat{\Sigma}^{(2)}$ at finite temperatures. Here, we concentrate our discussion on the diagonal part of $\hat{\Sigma}^{(2)}$, called $\delta \Sigma ''=\Sigma^{(2)}_{\rm{diag}}(T)-\Sigma^{(2)}_{\rm{diag}}(T=0)$, % in Eq.~\eqref{ansatzselfenergyeegraphene},
which enters the amplitude and frequency of the oscillation, see Eq. \eqref{LKformulaeeWWgraphene}, and was introduced in the ansatz for the self energy \eqref{ansatzselfenergyeegraphene}.
 We have found that 
 %\begin{widetext}
\begin{eqnarray}
\label{inelasticdampingk}
&&\delta \Sigma''({\bf{k}},\omega_n)=\nonumber\\&&-\frac{\alpha^2 \pi \ln 2}{12} \omega_n \left( \frac{T^2}{\omega_n^2+v_F^2 k^2}+ \mathcal{O}\left\{ \left(\frac{T^2}{\omega_n^2+v_F^2 k^2}  \right)^2\right\} \right),\ \ \ \ \ \ \
\end{eqnarray}
%\end{widetext}
\\
in an expansion of the integral \eqref{selfenergysecondorderexact} in $T^2/(\omega_n^2+v_F^2 k^2)$.
\\

%%%%%%%%%%%%%%%%%%%%%%%%%%%%%%%%%%%%
%Discussion
%%%%%%%%%%%%%%%%%%%%%%%%%%%%%%%%%%%%
\section{Discussion}
\label{ch:comparison}
In this Section we will give an overview of our main results and compare our findings for graphene to the two-dimensional electron gas. The derivation of the LK-formula for the two-dimensional Fermi gas was performed in Ref. \cite{Mirlin2006}.

\subsection{The Lifshitz-Kosevich formula in a clean system}

In Table \ref{tab:LKformulaclean} we contrast the Lifshitz-Kosevich formula for the 2DEG \cite{Mirlin2006} and for graphene without disorder or interaction effects.
We observe two peculiarities: (a) due to the restricted sum in Eq.~\eqref{eq:oscclean} the oscillations completely die as soon as $\mu<\pi T$; (b) the effective oscillation frequency is not only a geometric quantity any more but instead also depends on temperature itself via the dependence upon $\omega_n$. 
These differences become pronounced if we approach the Dirac point, meaning if temperature becomes comparable to the chemical potential. However, in the Fermi liquid regime of graphene, i.e., at $\mu\gg T$, the formula for a standard 2DEG is reproduced, albeit with the difference due to the differing spacing of Landau levels and the linear density of states.
\subsection{The Lifshitz-Kosevich formula in a disordered system}
In Table \ref{tab:LKformuladisorder} we contrast the LK-formula with disorder for graphene and for the 2DEG \cite{Mirlin2006}. Again, we see, that in contrast to the oscillations in the 2DEG, the oscillations in graphene die above a characteristic temperature $T_\textup{osc}$. This temperature depends on the strength $\alpha_{\textup{dis}}$ of the scattering potential and is given by $\pi T_\textup{osc}=\frac{|\mu|}{1+\alpha_{\textup{dis}}\frac{\pi}{2}}$. Disorder in graphene, unlike in the 2DEG, also affects the oscillation frequency.\\
In Table \ref{tab:LKformuladisorder} we also specify the Dingle temperature for the 2DEG as well as for graphene. We see, that in graphene, the Dingle temperature also depends on temperature, meaning there is an additional damping term due to the coupling of temperature and impurity scattering, which does not occur in the 2DEG. In Table \ref{tab:LKformuladisorder} we also give an account of the Dingle temperature for zero temperature, $T_D(T=0)$.
\begin{widetext}
\begin{table}
\caption{The LK-formula in a clean system}
\label{tab:LKformulaclean}
\vspace{1mm}
\begin{tabular}{ll}
\vspace{4mm}
\textbf{2DEG}   &\quad $\tilde{\Omega}_{osc}=4\nu\omega_c T\sum_{l=1}^{\infty}\sum_{\omega_n>0}^{\infty}\frac{(-1)^l}{l} e^{-\frac{2\pi l\omega_n}{\omega_c}}\cos\left(\frac{2\pi l \mu}{eB}\right)=4\nu\omega_c T\sum_{l=1}^{\infty}\frac{(-1)^l}{l} \frac{1}{\sinh\left(\frac{2\pi^2 l T}{\omega_c}\right)}\cos\left(\frac{2\pi l \mu}{eB}\right)$ \\
\vspace{5mm}
 \textbf{Graphene}  &\quad $\tilde{\Omega}_{osc}=\frac{2T\omega_c^2L^2}{\pi v_F^2}\sum_{l=1}^\infty \sum_{\omega_n>0}^{|\mu|}\frac{1}{l}e^{-\frac{4\pi l\omega_n |\mu|}{\omega_c^2}} \cos \left(\frac{2\pi l(\mu^2-\omega_n^2)}{\omega_c^2}\right)$\\
 \quad$\mu\gg T$ &\quad $\tilde{\Omega}_{\rm{osc}}\approx\frac{4T\omega_c^2L^2}{\pi v_F^2}\sum_{l=1}^\infty  \frac{1}{l \sinh \left(\frac{4\pi^2 l T |\mu|}{\omega_c^2} \right)} \cos \left(\frac{2\pi l \mu^2}{\omega_c^2} \right)$\\
 \end{tabular}\newline
\end{table}
\begin{table}
\caption{The LK-formula in a disordered system}
\label{tab:LKformuladisorder}
\vspace{1mm}
\begin{tabular}{ll}
\vspace{5mm}
\textbf{2DEG} & $\quad\tilde{\Omega}_{osc}=4\nu\omega_c T\sum_{l=1}^{\infty}\sum_{\omega_n>0}\frac{(-1)^l}{l} e^{-\frac{2\pi l}{\omega_c}(\omega_n+\frac{1}{2\tau})}\cos\left(\frac{2\pi l \mu}{eB}\right)$ \\
\vspace{3mm}
& $\quad T_D=\frac{1}{2\pi\tau}$\\
\vspace{3mm}
 \textbf{Graphene} & $\quad\tilde{\Omega}_{osc}=-\frac{2T\omega_c^2L^2}{\pi v_F^2}\sum_{l=1}^{\infty}\frac{1}{l}\sum_{\omega_n>0}^{\frac{|\mu|}{1+\alpha\frac{\pi}{2}}}e^{-\frac{4\pi l}{\omega_c^2}\left((\mu^2-\omega_n^2)(\pi\alpha-2\phi\alpha)+|\mu|\omega_n(1+2\alpha \Gamma)\right)}\cos\left(\frac{2\pi l}{\omega_c^2}\left( (\mu^2-\omega_n^2)(1+2\alpha \Gamma)-4\alpha|\mu|\omega_n(\pi-2\phi)\right)\right)$\\
 \vspace{3mm}
 &$\quad T_D=\omega_n\left(\frac{(\mu^2-\omega_n^2)}{|\mu|\omega_n}(\pi\alpha-2\phi\alpha)+2\alpha \Gamma\right)$\\
 \vspace{3mm}
&$\quad T_D(T=0)=|\mu| \pi \alpha$\\
 \end{tabular}
\end{table}
\end{widetext}

%%%%%%%%%%%%%%%%%%%%%%%%%%%%%%%%%%%%%%%%%%
\subsection{The effect of electron-electron interactions on the Lifshitz-Kosevich formula}

In Table \ref{tab:LKformulaeeWW} we contrast the LK-formula for weak interactions in graphene and in the 2DEG \cite{Mirlin2006}. One of the main features of the 2DEG is that inelastic processes on the first Matsubara mode do not lead to an additional damping Dingle temperature. \cite{Mirlin2006} This implies that interaction effects can fully be absorbed in renormalization factors.\\
The situation is different in graphene. Here, both inelastic effects as well as renormalization effects influence the amplitude. The inelastic effects are expressed by $\delta \Sigma'$ and $\delta \Sigma''$. The latter is given by (cf. Eq. \eqref{inelasticdampingk})
\begin{eqnarray}
\label{inelasticdamping}
&&\delta \Sigma''(x_0^{(0)},i\omega_n)=\nonumber\\&-&\frac{\alpha^2 \pi \ln 2}{12} \omega_n \Bigg( \frac{T^2}{(\omega_n^2+\mu^2)(1-Z^{-2})+2i\omega_nZ^{-2}\mu}\nonumber\\&& \quad+\mathcal{O}\left\{ \left(\frac{T^2}{(\omega_n^2+\mu^2)(1-Z^{-2})+2i\omega_nZ^{-2}\mu}  \right)^2\right\} \Bigg).\ \ \ \ \ \ \ 
\end{eqnarray}
However, unlike in the case of disorder, the inelastic effects vanish for zero temperature, as one should expect.
\begin{widetext}
\begin{table}[h]
\caption{The LK-formula with electron-electron interactions}
\label{tab:LKformulaeeWW}
\vspace{1mm}
\begin{tabular}{ll}
\vspace{3mm}
\textbf{2DEG} & 
$\quad\tilde{\Omega}_{osc}=-2DT\sum_{\omega_n>0} \sum_{l=1}^{\infty}\frac{1}{l} e^{-\frac{2\pi l(\omega_n(1+\alpha_0)-\delta\Sigma'')}{\omega_c(1+\beta)}}\cos\left(\frac{2\pi l\tilde{\mu}}{\omega_c(1+\beta)}\right)$ \\
\vspace{3mm}
 \textbf{Graphene} &$\quad\tilde{\Omega}_{osc}=2DT \sum_{l=1}^\infty\frac{1}{l}\sum_{\omega_n>0}^{\approx |\mu|}e^{-\frac{4\pi l}{\omega_c^2Z_{v_F}^2}\left(\omega_n\mu Z^{-2}+(\Im\{\delta\Sigma''\}+\Re\{\delta\Sigma'\})\omega_nZ^{-1}+(\Im\{\delta\Sigma'\}-\Re\{\delta\Sigma''\})\mu Z^{-1}\right)}$\\
 \vspace{5mm}
 &$ \quad \quad \quad \ \times\cos\left(\frac{2\pi l}{\omega_c^2Z_{v_F}^2}\Big((-\omega_n^2+\mu^2) Z^{-2}+2\big(\Re\{\delta\Sigma''\}-\Im\{\delta\Sigma'\}\big)\omega_nZ^{-1}\right.+2\big(\Im\{\delta\Sigma''\}+\Re\{\delta\Sigma'\}\big)\mu Z^{-1}\Big)\bigg)$\\
 \end{tabular}
\end{table}
\end{widetext}
Another interesting property is that the dominant damping term
\begin{eqnarray}\label{dominantdamping}
e^{-\frac{4\pi \omega_n \mu l}{\omega_c^2Z_{v_F}^2 Z^{2}}}= e^{-\frac{4\pi \omega_n \mu l}{(\omega^R_c)^2}}
\end{eqnarray}
is fully accounted for by renormalizations of the Fermi velocity or cyclotron frequency. So this result is in agreement with a recent analysis, Ref.~\cite{Elias2011}, carried out for Shubnikov-de Haas oscillations.

\section{Conclusion}
We derived a full quantitative expression which describes the de Haas - van Alphen oscillations in clean, disordered, and interacting Dirac systems. In the Fermi liquid regime of the system we reproduce the standard Lifshitz-Kosevich formula for the two-dimensional electron gas, despite the differences due to a different dispersion, which yields $m=\frac{\mu}{v_F^2}$ for the cyclotron mass in Dirac systems. However, when approaching the Dirac point, we found two new features in the clean system: first, the frequency is not only a geometric quantity but instead it also depends on temperature itself and second, the oscillations completely die as soon as $\mu<\pi T$. In the case of a disordered system, we find that disorder also affects the oscillation frequency and that there is an additional damping term due to the coupling of temperature and impurity scattering.\\
Most interesting is the effect of electron-electron interactions on the oscillation amplitude. 
We find that electron-electron interactions in Dirac systems damp the oscillations in two ways. They renormalize the Fermi velocity and to two-loop order they lead to an additional damping factor. This damping factor is absent in the more standard two-dimensional electron gas. In an extension of this work, it would be very interesting to see whether our results for the LK-formula would also survive for the case of Shubnikov-de Haas oscillations.  \\

%%%%%%%%%%%%%%%%%%%%%%%%%%%%%%%%%%%%
% Acknowledgement
%%%%%%%%%%%%%%%%%%%%%%%%%%%%%%%%%%%%

%\section{Acknowledgement}
{\it {Acknowledgement:}}
We acknowledge discussions with A. Rosch and I. Herbut as well as financial support from the DFG FR 2627/3-1. This work is part of the D-ITP consortium, a program of the Netherlands Organisation for Scientific Research (NWO) that is funded by the Dutch Ministry of Education, Culture and Science (OCW). 

%%%%%%%%%%%%%%%%%%%%%%%%%%%%%%%%%%%%
%%%%%%%%%%%%%%%%%%%%%%%%%%%%%%%%%%%%
% Appendix
%%%%%%%%%%%%%%%%%%%%%%%%%%%%%%%%%%%%
%%%%%%%%%%%%%%%%%%%%%%%%%%%%%%%%%%%%

\begin{appendix}
%%%%%%%%%%%%%%%%%%%%%%%%%%%%%%%%%%%%
% The oscillatory potential for the clean limit
%%%%%%%%%%%%%%%%%%%%%%%%%%%%%%%%%%%%

\section{The oscillatory integral for clean graphene}
\label{app:clean}
\begin{figure}[h]
\includegraphics[width=0.6\linewidth]{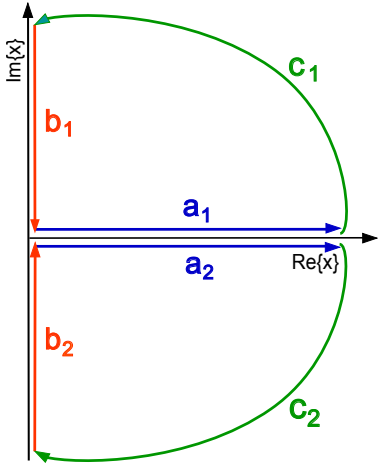}
\caption{Integration path: we use the integration path 1 in the upper complex half-plane to evaluate the integral $ \frac{e^{i2\pi lx}}{(i \omega_n+\mu)^2-\omega_c^2x}$ and path 2 in the lower complex half-plane for the integral$ \frac{e^{-i2\pi lx}}{(i \omega_n+\mu)^2-\omega_c^2x}$ such that the paths $c_{1,2}$ vanish for $x\rightarrow\infty$. Thus the original integral along the real axis, paths $a_{1,2}$ , can be written as an integral along the imaginary axis, paths $b_{1,2}$, plus summation over residues.}
\label{fig:integrationpath}
\end{figure}

The integral of Eq. \eqref{omegacleanbeforerestheorem} can be evaluated using residue theorem. The used integration path is described in Figure \ref{fig:integrationpath}.
\begin{widetext}
\begin{eqnarray}
\nonumber \int_{0}^\infty {\rm d}x \frac{\sin(2\pi lx)}{(i \omega_n+\mu)^2-\omega_c^2x}&=&\nonumber\frac{1}{2i} \int_{0}^\infty  {\rm d}x \frac{e^{i2\pi lx}-e^{-i2\pi lx}}{(i \omega_n+\mu)^2-\omega_c^2x}\\
&=&\nonumber \frac{1}{2}\int_0^{\infty}{\rm d}x \ e^{-2\pi lx}\left( \frac{1}{(i\omega_n+\mu)^2-i\omega_c^2x)}+\frac{1}{(i\omega_n+\mu)^2+i\omega_c^2x)}\right)\\&-&\frac{\pi}{\omega_c^2}\Theta(\mu^2-\omega_n^2)\left[e^{\frac{-4\pi l \omega_n\mu}{\omega_c^2}}e^{\frac{i2\pi l(\mu^2-\omega_n^2)}{\omega_c^2}}\Theta(\omega_n\mu)+e^{\frac{4\pi l \omega_n\mu}{\omega_c^2}}e^{\frac{-i2\pi l(\mu^2-\omega_n^2)}{\omega_c^2}}\Theta(-\omega_n\mu)\right]
\end{eqnarray}
If we use $\sum_{\omega_n}F(\omega_n)= \sum_{\omega_n>0}\left(F(\omega_n)+F(-\omega_n)\right)$ we get the following expression,
\begin{eqnarray}\label{afterres}
\sum_{\omega_n}\int_{0}^\infty {\rm d}x \frac{\sin(2\pi lx)}{(i \omega_n+\mu)^2-\omega_c^2x}&=&\nonumber\sum_{\omega_n>0}(\mu^2-\omega_n^2)\int_{0}^\infty {\rm d}x \ e^{-2\pi lx}\left(\frac{1}{(\mu^2-\omega_n^2)^2+(2\omega_n\mu-\omega_c^2x)}+\frac{1}{(\mu^2-\omega_n^2)^2+(2\omega_n\mu+\omega_c^2x)}\right)\\
&-&\frac{\pi}{\omega_c^2}\sum_{\omega_n>0}e^{\frac{-4\pi l\omega_n|\mu|}{\omega_c^2}}2\cos\left(\frac{2\pi l}{\omega_c^2}(\mu^2-\omega_c^2)\right)\Theta(\mu^2-\omega_n^2).
\end{eqnarray}
\end{widetext}
Importantly, the first term on the right-hand-side in Eq. \eqref{afterres} does not contribute an oscillatory term to the thermodynamic potential.
%%%%%%%%%%%%%%%%%%%%%%%%%%%%%%%%%%%%
% The disorder induced self-energy}
%%%%%%%%%%%%%%%%%%%%%%%%%%%%%%%%%%%%

\section{The disorder induced self-energy}
\label{app:selfenergydis}
It is well known that within the self-consistent Born approximation (SCBA) there is a difference between the self-energy on the different sublattices~\cite{BriskotMirlin}. The defining equation reads
\begin{equation}
\Sigma_{\textup{dis}}^{a(b)}=\alpha_{\textup{dis}} \omega_c^2\sum_{m}\frac{i\omega_n+\mu-\Sigma_{\textup{dis}}^{b(a)}}{(i\omega_n+\mu-\Sigma_{\textup{dis}}^a)(i\omega_n+\mu-\Sigma_{\textup{dis}}^b)-\omega_c^2m},
\end{equation}
where $\alpha_{\textup{dis}}$ is a dimensionless parameter characterizing the strength of the disorder potential.\\
In the following we assume that sufficiently high-lying Landau levels are populated such that the asymmetry between the sublattices is irrelevant. We then have,
\begin{equation}
\Sigma_{\textup{dis}}=\alpha_{\textup{dis}} \omega_c^2\sum_{m=0}^\infty \frac{i\omega_n +\mu-\Sigma_{\textup{dis}}}{(i\omega_n + \mu -\Sigma_{\textup{dis}})^2-\omega_c^2 m}\;.
\end{equation}
We use the Poisson summation formula, Eq.~\eqref{Poisson}, to dualize the sum over Landau levels making the expression more amenable to approximations for weak fields.\\
\begin{eqnarray}
\label{fullselfenergydisorder}
\Sigma_{\textup{dis}} &=&\alpha_{\textup{dis}}\omega_c^2\sum\limits_{l=-\infty}^{\infty} \int_{0}^{\frac{v_F\Lambda}{\omega_c}}  {\rm d}x \frac{i\omega_n+\mu-\Sigma_{\textup{dis}}}{(i\omega_n+\mu-\Sigma_{\textup{dis}})^2-\omega_c^2x} e^{i2\pi lx}\nonumber \\ 
&=&\alpha_{\textup{dis}}\int_{0}^{v_F^2 \Lambda^2} {\rm d}x \frac{i\omega_n+\mu-\Sigma_{\textup{dis}}}{(i\omega_n+\mu-\Sigma_{\textup{dis}})^2-x} \nonumber \\ &+& \alpha_{\textup{dis}} \sum_{l=1}^\infty \int_0^{v_F^2\Lambda^2} {\rm d}x \frac{(i\omega_n+\mu-\Sigma_{\textup{dis}})}{(i\omega_n+\mu-\Sigma_{\textup{dis}})^2- x}\cos \left( \frac{2\pi l x}{\omega_c^2}\right)\nonumber \\
&=& \Sigma_{\textup{dis}}^0+\Sigma_{\textup{dis}}^{osc}
\end{eqnarray}
Here we introduced a cutoff $\Lambda$, restricting the analysis to the regime where the dispersion is linear.
The first term $\Sigma_{\textup{dis}}^0$ is the $l=0$ term and corresponds to the standard expression of the SCBA in a system without magnetic field. The second part $\Sigma_{\textup{dis}}^{osc}$ describes the oscillations of the self-energy due to the magnetic field. Performing the integration yields 
\begin{equation}
\Sigma_{\textup{dis}}^0 =-\alpha_{\textup{dis}} (i \omega_n+\mu-\Sigma_{\textup{dis}})\ln \frac{v_F^2 \Lambda^2-(i \omega_n+\mu-\Sigma_{\textup{dis}})^2}{-(i \omega_n+\mu-\Sigma_{\textup{dis}})^2}.
\end{equation}
This expression can be solved self consistently to leading order in $\alpha_{\textup{dis}}$ and we obtain,
\begin{equation}
\Sigma_{\textup{dis}}^0 =-\alpha_{\textup{dis}} (i \omega_n+\mu)\ln \frac{v_F^2 \Lambda^2-(i \omega_n+\mu)^2}{-(i \omega_n+\mu)^2}.
\end{equation}
The oscillatory part, $\Sigma_{\textup{dis}}^{osc}$, is treated by using residue theorem,
\begin{widetext}
\begin{eqnarray}
\Sigma_{\textup{dis}}^{osc} &=&\alpha_{\textup{dis}} ( i\omega_n+\mu-\Sigma_{\textup{dis}})\sum_{l=1}^\infty \int_0^\infty {\rm d}x \frac{e^{ i\frac{2\pi l x}{\omega_c^2}}+e^{-i\frac{2\pi l x}{\omega_c^2}}}{(i\omega_n+\mu-\Sigma_{\textup{dis}})^2- x}\nonumber \\
 &= &\nonumber\alpha_{\textup{dis}} \left (i \omega_n+\mu-\Sigma_{\textup{dis}} \right)2 \pi i \Theta(a) \sum_{l=1}^\infty  e^{-\frac{2\pi}{\omega_c^2}|b|l}(e^{i\frac{2\pi}{\omega_c^2}al}\Theta(b)-e^{-i\frac{2\pi}{\omega_c^2}al}\Theta(-b))\\&+&
\alpha_{\textup{dis}} \left (i \omega_n+\mu-\Sigma_{\textup{dis}} \right)\sum_{l=1}^\infty \int_0^\infty {\rm d}x \left(\frac{e^{-\frac{2\pi}{\omega_c^2}xl}}{a+i (b-x)}-\frac{e^{-\frac{2\pi}{\omega_c^2}xl}}{a+i (b+x)}\right)
\end{eqnarray}
\end{widetext}
with \ \ $a=(\mu-\Sigma_{\textup{dis}}^{'})^2-(\omega_n-\Sigma_{\textup{dis}}^{''})^2$ and $b=2(\omega_n-\Sigma_{\textup{dis}}^{''})(\mu-\Sigma_{\textup{dis}}^{'})$. $\Sigma_{\textup{dis}}^{'}$ is the real part of $\Sigma_{\textup{dis}}$ and $\Sigma_{\textup{dis}}^{''}$ is its imaginary part. 
The first term stems from the residue. Whether the pole is located inside or outside the integration contour depends on the sign of $a$ and $b$. The second term stems from integration along the imaginary axis and is a non-oscillatory correction term. The integration contours used here are plotted in Figure \ref{fig:integrationpath}. 
%The integral over $x$ is solved by applying integration by parts iteratively. The sum over $l$ can here also be solved by mathematica.
%\begin{equation}
%\Sigma_{\textup{dis}}^{osc} &=& \nonumber
%\alpha_{\textup{dis}} \left (i \omega_n+\mu-\Sigma_{\textup{dis}} \right)\pi i \Theta(a)\left( \frac{\Theta(b)}{e^{\frac{2\pi}{\omega_c^2}(|b|-ia)}-1}-%\frac{\Theta(-b))}{e^{\frac{2\pi}{\omega_c^2}(|b|+ia)}-1}\right)\\&+&
%\alpha_{\textup{dis}} \left (i \omega_n+\mu-\Sigma_{\textup{dis}} \right) 2i\sum_{l=1}^\infty \sum_{n=1}^\infty\left( \frac{\omega_c^2}{2\pi l(a+ib)}\right)^{2n}(2n+1)!(-1)^{n+1}\nonumber\\
%&=&\alpha_{\textup{dis}} \left (i \omega_n+\mu-\Sigma_{\textup{dis}} \right)\pi i \Theta(a)\left( \frac{\Theta(b)}{e^{\frac{2\pi}{\omega_c^2}(|b|-ia)}-1}-\frac{\Theta(-b))}{e^{\frac{2\pi}{\omega_c^2}(|b|+ia)}-1}\right)\\&+&
%\alpha_{\textup{dis}} \left (i \omega_n+\mu-\Sigma_{\textup{dis}} \right) 2i\sum_{l=1}^\infty \sum_{n=1}^\infty\left( \frac{\omega_c^2}{2\pi l(a+ib)}\right)^{2n}(2n+1)!(-1)^{n+1}\nonumber\\
%\end{equation}
In the regime $T\gtrsim\omega_c$ both terms in $\Sigma_{\textup{dis}}^{osc}$ are suppressed exponentially due to the factors $\exp(-\frac{2\pi}{\omega_c^2}|b|l)$ and $\exp(-\frac{2\pi}{\omega_c^2}xl)$, respectively. Thus, in this regime, $\Sigma_{\textup{dis}}^{osc}$ can be neglected, and the self-energy can be well-approximated as,
\begin{equation}
\Sigma_{\textup{dis}}\approx\Sigma_{\textup{dis}}^{0}.
\end{equation}

%%%%%%%%%%%%%%%%%%%%%%%%%%%%%%%%%%%%
% Oscillatory grand potential with disorder
%%%%%%%%%%%%%%%%%%%%%%%%%%%%%%%%%%%%

\section{Oscillatory grand potential with disorder}
\label{app:dis}
In this section we will compute the grand potential for graphene with disorder, Eq. \eqref{oscintdis},
%In the presence of disorder the calculation of the oscillatory integral is analogue to that for clean graphene with the self energy entering the Greenfunction.
\begin{equation}
\tilde{\Omega}=-\frac{T\omega_c^4L^2}{\pi^2 v_F^2}\sum_{\omega_n} \sum_{l=1}^{\infty}\frac{1}{l}\int_{0}^\infty  {\rm d}x \frac{\sin(2\pi lx)}{\left[ i \omega_n+\mu+\Sigma_{\textup{dis}}\right]^2-\omega_c^2x}\nonumber\\
\end{equation}
The computation is analogue to the one for the clean system which we perform in Appendix \ref{app:clean}. However, it is more tricky to find the location of the pole, which is needed to evaluate this integral using residue theorem. The pole is given by
\begin{eqnarray}
x\nonumber&=&\frac{1}{\omega_c^2}[-i\omega_n+\mu+\Sigma_{\textup{dis}}]^2\\ \vspace{3mm}&=&\frac{1}{\omega_c^2}(i\omega_n-\mu)^2\left(1+\alpha_{\textup{dis}}\ln\left[\frac{(v_F\Lambda)^2}{-(i\omega_n-\mu)^2}\right]\right)^2.\ \ \ 
\end{eqnarray}\\
We expand the pole to linear order in $\alpha_{\textup{dis}}$ as we are interested in a weak disorder potential.\\
\begin{eqnarray*}
&\omega_c^2x\nonumber& \ \approx(\mu^2-\omega_n^2)\left(1+\alpha_{\textup{dis}}\ln\left[\frac{v_F\Lambda}{r}\right]^2\right)-2\omega_n\mu\alpha_{\textup{dis}}(\pi-2\phi)\\&-&i\left(2\omega_n\mu\left(1+\alpha_{\textup{dis}}\ln\left[\frac{v_F\Lambda}{r}\right]^2\right)+(\mu^2-\omega_n^2)\alpha_{\textup{dis}}(\pi-2\phi)\right)
\end{eqnarray*}
with $r=\omega_n^2+\mu^2$ and $\phi=\arctan\left(\frac{\omega_n}{\mu}\right)$.
The imaginary part is always smaller than zero. 
In order to find the zero-crossing of the real part we use the ansatz $\mu=\omega_n+\delta\mu$ and expand the real part up to first oder in $\delta\mu$. We find the zero-crossing at $\mu=\omega_n(1+\alpha_{\textup{dis}}\frac{\pi}{2})$ such that only frequencies $|\omega_n|<\frac{|\mu|}{1+\alpha_{\textup{dis}}\frac{\pi}{2}}$ contribute to the oscillations.
We find 
\begin{widetext}
\begin{eqnarray}
\nonumber \tilde{\Omega}&=&-\frac{T\omega_c^4L^2}{\pi^2 v_F^2}\sum_{l=1}^{\infty}\frac{1}{l}\sum_{\omega_n>0}\int_0^{\infty}\textup{d}x \ e^{-2\pi lx}\left((\mu+\Sigma'_{\textup{dis}})^2-(\omega_n-\Sigma''_{\textup{dis}})^2\right)\\ \nonumber &\times&\Bigg(\frac{1}{\left((\mu+\Sigma'_{\textup{dis}})^2-(\omega_n-\Sigma''_{\textup{dis}})^2\right)^2 +(2(\mu+\Sigma'_{\textup{dis}})(\omega_n-\Sigma''_{\textup{dis}}
)+\omega_c^2 x)}\nonumber\\&+&\frac{1}{\left((\mu+\Sigma')^2-(\omega_n-\Sigma'')^2\right)^2+(2(\mu+\Sigma')(\omega_n-\Sigma'')-\omega_c^2x)}\Bigg)\nonumber\\&+&\frac{2T\omega_c^2L^2}{\pi v_F^2}\sum_{l=1}^{\infty}\frac{1}{l}\sum_{\omega_n>0}^{\frac{|\mu|}{1+\alpha_{dis} \frac{\pi}{2}}}e^{-\frac{4\pi l}{\omega_c^2}\left((\mu^2-\omega_n^2)(\pi\alpha-2\phi\alpha_{dis})+|\mu|\omega_n(1+2\alpha_{dis} \Gamma)\right)}\cos\left(\frac{2\pi l}{\omega_c^2}\left( (\mu^2-\omega_n^2)(1+2\alpha_{dis} \Gamma)+4\alpha_{dis}|\mu|\omega_n(2\phi-\pi)\right)\right)\nonumber
\end{eqnarray}
\begin{eqnarray}
\end{eqnarray}
\end{widetext} 
where $\Sigma'_{\textup{dis}}$ is the real part of the self-energy and $\Sigma''_{\textup{dis}}$ its imaginary part. $\Gamma$ is defined in \eqref{eq:Gamma}.
Here, again only the last term is oscillatory.

%%%%%%%%%%%%%%%%%%%%%%%%%%%%%%%%%%%%
% The interaction induced self-energy
%%%%%%%%%%%%%%%%%%%%%%%%%%%%%%%%%%%%

\section{The interaction induced self-energy}

 \label{app:selfenergyee}
In this section we will compute the interaction induced self-energy $\hat{\Sigma}_{ee}$ for graphene. The generic expression for the self-energy within RPA reads,
\begin{equation}
\hat{\Sigma}_{ee}({\bf{k}},i \omega_n)=-T\sum_{\omega_n'}\int \frac{d^2k'}{(2\pi)^2} V({\bf{k}}-{\bf{k}}',i\omega_n-i\omega_n')\hat{G}_0({\bf{k}}',i\omega_n')\;.
\end{equation}
The free electron Green function for graphene is given by
\begin{equation}
\hat{G}_0({\bf{k}},i\omega_n)=\frac{-i\omega_n\mathbb{1}-v_F {\bf{k}} \hat{\boldsymbol\sigma}}{\omega_n^2+v_F^2 k^2}
\end{equation}
and the Coulomb interaction $V({\bf{k}},i\omega_n)$ in the RPA approximation is given by
\begin{equation}
V({\bf{k}},i\omega_n)=\frac{2\pi \alpha v_F}{|{\bf{k}}|+\frac{2\pi \alpha v_F}{4}\frac{k^2}{4\sqrt{v_F^2 k^2+\omega_n^2}}} \;,
\end{equation}
with $\alpha=e^2/(\varepsilon v_F)$ ($\varepsilon$ corresponds to the dielectric constant) being graphene's dimensionless fine structure constant.
In the following we will only work to two-loop accuracy and consequently expand the dressed Coulomb interaction to quadratic order yielding
\begin{equation}
V({\bf{k}},i\omega_n) = \frac{2\pi \alpha v_F}{k}-\frac{(2\pi)^2 \alpha^2 v_F^2}{4 \sqrt{v_F^2 k^2+\omega_n^2}}+\mathcal{O}(\alpha^3)\;.
\end{equation}
We decompose the self-energy into a first and a second order part according to
\begin{equation}
\hat{\Sigma}_{ee}({\bf{k}},i\omega_n)\approx \hat{\Sigma}^{(1)}({\bf{k}},i\omega_n)+\hat{\Sigma}^{(2)}({\bf{k}},i\omega_n)
\end{equation}
where $\hat{\Sigma}^{(1)}({\bf{k}},i\omega_n)$ is linear in $\alpha$, while $\hat{\Sigma}^{(2)}({\bf{k}},i\omega_n)$ is quadratic in $\alpha$.
They read
\begin{widetext}
\begin{equation}
\hat{\Sigma}^{(1)}({\bf{k}},i\omega_n)=2\pi \alpha v_F T \sum_{\omega_n'}\frac{d^2 k'}{(2\pi)^2} \frac{1}{|{\bf{k}}-{\bf{k}}'|}\frac{i\omega_n' \mathbb{1}+ v_F {\bf{k}}' \hat{\boldsymbol\sigma} }{\omega_n'^2+v_F^2 k'^2}
\end{equation}
and 
\begin{equation}
\hat{\Sigma}^{(2)}({\bf{k}},i\omega_n)=-\frac{(2\pi)^2 \alpha^2 v_F^2}{4}T\sum_{\omega_n'} \int \frac{d^2 k'}{(2\pi)^2} \frac{1}{\sqrt{v_F^2({\bf{k}}-{\bf{k}}')^2+(\omega_n-\omega_n')^2}} \frac{i\omega_n' \mathbb{1}+ v_F {\bf{k}}' \hat{\boldsymbol\sigma} }{\omega_n'^2+v_F^2 k'^2}.
\end{equation}

We see that the imaginary part is strictly diagonal, while the real part is off-diagonal. This motivates ansatz \eqref{ansatzselfenergyeegraphene} for the self-energy.\\
We start with the calculation of $\hat{\Sigma}^{(1)}$. From symmetry we observe that the diagonal part vanishes and only the off-diagonal part survives:
\begin{equation}
\hat{\Sigma}^{(1)}({\bf{k}},i\omega_n)=2\pi \alpha v_F T \sum_{\omega_n'}\frac{d^2 k'}{(2\pi)^2} \frac{1}{|{\bf{k}}-{\bf{k}}'|}\frac{ v_F {\bf{k}}' \hat{\boldsymbol\sigma} }{\omega_n'^2+v_F^2 k'^2}
\end{equation}
We apply an integral identity,
\begin{align}
\label{intidentity}
\int_{-\infty}^\infty\frac{\textup{d}x}{\pi}\frac{1}{a^2+x^2}=\frac{1}{a},
\end{align}
and rescale $\bf k'$ and $x$ with $v_F$, and obtain,
\begin{equation*}
\hat{\Sigma}^{(1)} ({\bf  k},i \omega_n)=2\pi\alpha T\sum_{\omega'_n} \int \frac{d^2 k'}{(2\pi)^2}\int\frac{dx}{\pi}\frac{1}{({\bf k'}-v_F{\bf k})^2+x^2}\frac{{\bf k'}\hat{\boldsymbol{\sigma}}}{\omega_n^{'2}+k^{'2}}.
\end{equation*} 
We use the Feynman parameter and write,
\begin{eqnarray*}
\hat{\Sigma}^{(1)} ({\bf  k},i \omega_n)&=&2\pi \alpha T\sum_{\omega'_n} \int \frac{d^2 k'}{(2\pi)^2}\int\frac{dx}{\pi}\int_0^1du\frac{{\bf k'}\hat{\boldsymbol{\sigma}}}{\left(u({\bf k'}-v_F{\bf k})^2+ux^2+(1-u)(\omega_n^{'2}+k^{'2})\right)^2}\nonumber\\
&=&2\pi\alpha T\sum_{\omega'_n} \int \frac{d^2 k'}{(2\pi)^2}\int\frac{dx}{\pi}\int_0^1du\frac{{\bf k'}\hat{\boldsymbol{\sigma}}}{\left(({\bf k'}-uv_F{\bf k})^2+ux^2+u(1-u)v_F^2k^{2}+(1-u)\omega_n^{'2})\right)^2}\nonumber\\
&=&2\pi \alpha T\sum_{\omega'_n} \int \frac{d^2 k'}{(2\pi)^2}\int\frac{dx}{\pi}\int_0^1du\frac{1}{\sqrt{u}}\frac{({\bf k'}+uv_F{\bf k})\hat{\boldsymbol{\sigma}}}{\left(k^{'2}+x^2+u(1-u)v_F^2k^{2}+(1-u)\omega_n^{'2})\right)^2}.
\end{eqnarray*} 
In the last step we shifted ${\bf k}\rightarrow{\bf k'}+uv_F{\bf k}$ and rescaled $x$ and $\omega'_n$. The $\bf k'$ integral over the first summand vanishes as the integrand is odd, the second summand contributes,
\begin{eqnarray}
\hat{\Sigma}^{(1)} ({\bf  k},i \omega_n)&=&-\frac{\alpha Tv_F{\bf k}\hat{\boldsymbol{\sigma}}}{2}\sum_{\omega'_n} \int\frac{dx}{\pi}\int_0^1du\sqrt{u}\left(\frac{1}{\Lambda^2+x^2+u(1-u)v_F^2k^{2}+(1-u)\omega_n^{'2}}\right.-\left.\frac{1}{x^2+u(1-u)v_F^2k^{2}+(1-u)\omega_n^{'2}}\right).\nonumber\\
&=&-\frac{\alpha Tv_F{\bf k}\hat{\boldsymbol{\sigma}}}{2}\sum_{\omega'_n}\int_0^1du\sqrt{u}\left(\frac{1}{\sqrt{\Lambda^2+u(1-u)v_F^2k^{2}+(1-u)\omega_n^{'2}}}-\frac{1}{\sqrt{u(1-u)v_F^2k^{2}+(1-u)\omega_n^{'2}}}\right).
\end{eqnarray} 
The sum over $\omega'_n$ can only be performed analytically in the limit $T\rightarrow0$ when one can transform the sum into an integral,
\begin{eqnarray}
\hat{\Sigma}^{(1)} ({\bf  k},i \omega_n)&=&-\frac{\alpha v_F{\bf k}\hat{\boldsymbol{\sigma}}}{4 \pi}\int d\omega'\int_0^1du\sqrt{\frac{u}{1-u}}\left(\frac{1}{\sqrt{\Lambda^2+u(1-u)v_F^2k^{2}+\omega^{'2}}}-\frac{1}{\sqrt{u(1-u)v_F^2k^{2}+\omega^{'2}}}\right).\nonumber\\
&=&-\frac{\alpha v_F{\bf k}\hat{\boldsymbol{\sigma}}}{4 \pi}\int_0^1du\sqrt{\frac{u}{1-u}}\ln\left(\frac{u(1-u)v_F^2k^{2}}{\Lambda^2+u(1-u)v_F^2k^{2}}\right)\nonumber\\
&\approx&-\frac{\alpha v_F{\bf k}\hat{\boldsymbol{\sigma}}}{4 \pi}\int_0^1du\sqrt{\frac{u}{1-u}}\ln\left(\frac{u(1-u)v_F^2k^{2}}{\Lambda^2}\right)\nonumber\\
&=&\frac{\alpha v_F{\bf k}\hat{\boldsymbol{\sigma}}}{4 }\ln\left(\frac{4\Lambda}{v_Fk}\right).
\end{eqnarray} 
Now we calculate $\hat{\Sigma}^{(2)}({\bf  k},i \omega_n)$. We again use the integral identity Eq. \eqref{intidentity} and write,
\begin{equation*}
\hat{\Sigma}^{(2)}({\bf{k}},i\omega_n)=-\frac{(2\pi)^2 \alpha^2 v_F^2}{4}T\sum_{\omega_n'} \int \frac{d^2 k'}{(2\pi)^2} \int \frac{dx}{\pi}\frac{1}{x^2+v_F^2({\bf{k}}-{\bf{k}}')^2+(\omega_n-\omega_n')^2} \frac{i\omega_n' \mathbb{1}+ v_F {\bf{k}}' \hat{\boldsymbol\sigma} }{\omega_n'^2+v_F^2 k'^2}\ \ .
\end{equation*}
Using the standard Feynman parameter we can rewrite it as
\begin{equation}
\label{selfenergysecondorderexact}
\hat{\Sigma}^{(2)}({\bf{k}},i\omega_n)=-\pi^2 \alpha^2 v_F^2T\sum_{\omega_n'} \int \frac{d^2 k'}{(2\pi)^2} \int \frac{dx}{\pi} \int_0^1 du \frac{i\omega_n' \mathbb{1}+ v_F {\bf{k}}' \hat{\boldsymbol\sigma}}{u x^2+v_F^2({\bf{k}}'-u{\bf{k}})^2+(\omega_n'-u\omega_n)^2+u(1-u)\Omega^2} 
\end{equation}
where $\Omega^2=v_F^2 k^2+\omega_n^2$.
In the following we analyze this expression in the zero temperature limit. We note, however, that we have also analyzed the finite temperature behavior of this expression numerically, given in Eq. \eqref{inelasticdampingk}. In the limit $T \to 0$ we can rewrite the expression after an appropriate shift as
\begin{equation*}
\hat{\Sigma}^{(2)}({\bf{k}},i\omega_n)=-\pi^2 \alpha^2 v_F^2 \int \frac{d \omega'}{2\pi} \int \frac{d^2 k'}{(2\pi)^2} \int \frac{dx}{\pi} \int_0^1 du \frac{(i\omega'+u i \omega_n )\mathbb{1}+ v_F ({\bf{k}}'+u{\bf{k}}) \hat{\boldsymbol\sigma}}{u x^2+v_F^2k'^2+\omega'^2+u(1-u)\Omega^2}.
\end{equation*} For symmetry reasons we can simplify the expression to yield
\begin{equation*}
\hat{\Sigma}^{(2)}({\bf{k}},i\omega_n)=-\pi^2 \alpha^2  ( i \omega_n \mathbb{1}+ v_F {\bf{k}} \hat{\boldsymbol\sigma} )\int \frac{d \omega'}{2\pi} \int \frac{d^2 k'}{(2\pi)^2} \int \frac{dx}{\pi} \int_0^1 du  \frac{\sqrt{u}}{x^2+k'^2+\omega'^2+u(1-u)\Omega^2}\ .\ \ 
\end{equation*}
The integrals over $k'$ with a cutoff $\Lambda$ and $\omega'$ are elementary and we obtain
\begin{equation*}
\hat{\Sigma}^{(2)}({\bf{k}},i\omega_n)=- \frac{\alpha^2}{8}  ( i \omega_n \mathbb{1}+ v_F {\bf{k}} \hat{\boldsymbol\sigma} )  \int_0^1 du \sqrt{u} \int dx\left(\frac{1}{\sqrt{x^2+u(1-u)\Omega^2}}\right.-\left.\frac{1}{\sqrt{x^2+u(1-u)\Omega^2+\Lambda^2}}\right)\ .
\end{equation*}
Integrating over $x$ leaves us with
\begin{equation*}
\hat{\Sigma}^{(2)}({\bf{k}},i\omega_n)=- \frac{\alpha^2}{8}  ( i \omega_n \mathbb{1}+ v_F {\bf{k}} \hat{\boldsymbol\sigma} )  \int_0^1 du \sqrt{u} \ln \frac{u(1-u)\Omega^2+\Lambda^2}{u(1-u)\Omega^2}\;,
\end{equation*}
which we can integrate to give
 \begin{eqnarray}
\hat{\Sigma}^{(2)}({\bf{k}},i\omega_n)&=& - \frac{\alpha^2}{8}  ( i \omega_n \mathbb{1}+ v_F {\bf{k}} \hat{\boldsymbol\sigma} )\int_0^1{\textup{d}u}\ \sqrt{u}\ln\left(\frac{u(1-u)\Omega^2+\Lambda^2}{u(1-u)\Omega^2}\right)\nonumber\\
 &=& \frac{\alpha^2}{24}( i \omega_n \mathbb{1}+ v_F {\bf{k}} \hat{\boldsymbol\sigma} )\Bigg[\frac{\sqrt{2} }{\Omega ^{\frac{3}{2}}} \Bigg(\left(-\Omega +\sqrt{4 \Lambda ^2+\Omega ^2}\right)^{3/2} \arctan\left[\frac{\sqrt{2\Omega}}{\sqrt{-\Omega
+\sqrt{4  \Lambda ^2+\Omega ^2}}}\right]\nonumber\\&+&\left(-\Omega-\sqrt{4 \Lambda ^2+\Omega ^2}\right)^{3/2}
\arctan\left[\frac{\sqrt{2\Omega}}{\sqrt{- \Omega -\sqrt{4  \Lambda ^2+\Omega ^2}}}\right]\Bigg)+2\ln\left[\frac{4 \Omega ^2}{
\Lambda ^2}\right]\Bigg]\;.
\end{eqnarray}
In the limit $\Omega^2 \ll \Lambda^2$ this reduces to 
 \begin{equation}
\hat{\Sigma}^{(2)}({\bf{k}},i\omega_n)=- \frac{\alpha^2}{2}  ( i \omega_n \mathbb{1}+ v_F {\bf{k}} \hat{\boldsymbol\sigma} ) \left(\frac{1}{6} \ln \left(\frac{\Lambda ^2 v_F^2}{k^2 v_F^2+\omega_n^2}\right)-\frac{1}{3}\ln (2)+\frac{5}{9}\right).\ 
\end{equation}
\end{widetext}

\end{appendix}

\end{document}